\documentclass[hyper,letterpaper,12pt]{article}
\usepackage{a4wide}
\usepackage{amsmath}
\usepackage[
      colorlinks=true,
      linkcolor=blue,
      urlcolor=blue,
      filecolor=black,
      citecolor=blue,
            ]{hyperref}
\usepackage{color}
\usepackage{graphicx}
\usepackage{amsfonts}
\usepackage{amssymb}
\usepackage{epsfig}
\usepackage{subfigure}
\usepackage{amsmath}
\usepackage{latexsym}
\usepackage{mathrsfs}

\begin{document}
\title{\bf \Large Holographic phase transition probed by non-local observables}

\author{\large
~~Xiao-Xiong Zeng$^{1,2}$\footnote{E-mail: xxzeng@itp.ac.cn}~,
~~Li-Fang Li$^3$\footnote{E-mail: lilf@itp.ac.cn}~
\date{\today}
\\
\\
\small $^1$ School of Material Science and Engineering, Chongqing Jiaotong University,\\
\small       Chongqing ~400074, China\\
\small $^2$State Key Laboratory of Theoretical Physics, Institute of Theoretical Physics,\\
\small Chinese Academy of Sciences, Beijing 100190,  China\\
\small $^3$ Center for Space Science and Applied Research, Chinese Academy of Sciences,\\
\small Beijing 100190, China\\}

\maketitle

\begin{abstract}
\normalsize From the viewpoint of holography, the phase structure of a 5-dimensional  Reissner-Nordstr\"{o}m-AdS  black hole is probed by the two point correlation function, Wilson loop, and  entanglement entropy. As the case of thermal entropy, we find for all the probes, the black hole undergos a Hawking-Page phase transition, a first order phase transition and a second  order phase transition successively  before it reaches to a stable phase. In addition, for these probes, we find the equal area law for the first order phase transition is valid always and the critical exponent of the heat capacity for the second order phase transition coincides with that of the mean field theory   regardless of the  size of the boundary region.
\end{abstract}

\newpage

\tableofcontents

\section{Introduction}

Phase transition is a ubiquitous phenomenon for garden-variety thermodynamic systems. Due to the pioneering work by Hawking \cite{Hawking2,Hawking3}, a black hole is also a thermodynamic system. Such a fact is further supported by AdS/CFT correspondence \cite{ads1,ads2,ads3}, where a black hole in the AdS bulk  is dual to a thermal system without gravity. So one can naturally expect a black hole can also undertake some interesting phase transitions as the general thermodynamic system. Actually it has been shown  that  a charged AdS black hole
 undergos
 a Hawking-Page phase transition \cite{Hawking4,Caihp}, which is interpreted as the confinement/deconfinement
phase transition in the dual gauge field theory \cite{Witten1}, and a Van der Waals-like phase transition before it reaches the stable state \cite{Chamblin}.  The Hawking-Page phase transition implies that the thermal AdS is unstable and it will transit to the stable Schwarzschild AdS black hole lastly.
The Van der Waals-like phase transition has been  observed till now in many circumstantces.
The first observation was contributed by \cite{Chamblin}  in the $T-S$ plane.
Specifically speaking,  in a fixed charge ensemble,
for a black hole endowed with small
charge, there is an unstable black hole interpolating between the stable small hole and stable large
 hole,  and the small stable hole will undertake a first order phase transition to the large stable hole as the temperature of the black hole reaches a critical temperature.
  As the charge increases to the critical charge, the small hole and the large hole merge into one and squeeze  out the unstable phase so that an inflection point emerges and the phase transition is  second order.
 When the charge exceeds the critical charge, the  black hole is always stable.
   Recently in the extended phase space, where the negative cosmological constant is treated as the  pressure
while its conjugate acts as the thermodynamical volume, the Van der Waals-like phase transition has also been observed in the $P-V$
plane \cite{Kubiznak,
Xu, Caipv, Hendi, Hennigar, Wei1, Mo}.  In addition, it was shown in \cite{Niu} that  the  Van der Waals-like phase transition also shows up in the  $Q-\Phi$ plane. Particularly, in the Gauss-Bonnet gravity, it is found that the Gauss-Bonnet  coupling parameter $\alpha$ also affects the phase structure of the space time, and in the $T-\alpha$ plane, a 5-dimensional neutral  Gauss-Bonnet  black hole also demonstrates the Van der Waals-like phase transition \cite{Cai}.

In this paper, we intend to probe the Hawking-Page phase transition and Van der Waals-like phase transition
appeared in a 5-dimensional  Reissner-Nordstr\"{o}m-AdS  black hole by  the geodesic length, minimal area surface, and minimal surface area in the bulk, which are dual to the non-local observables on the boundary theory by holography, namely the two point correlation function, Wilson loop, and  entanglement entropy, individually\footnote{Recently these non-local observables have been used to probe the non-equilibrium thermalization process, and it has been found that all of them have the same effect \cite{
Balasubramanian1,Balasubramanian2,GS,CK, Zeng2013, Zeng2014, Zeng20151}.}. In fact, there have been some similar works to probe the phase structure  by holographic entanglement entropy.
In \cite{Johnson}, the phase structure of entanglement entropy  is studied in the $T-S$ plane for both a fixed charge ensemble and a fixed chemical potential ensemble, and  it is found that the phase structure of entanglement entropy is similar to that of the thermal entropy. In particular,  the entanglement entropy is found to demonstrate the same second order phase transition at the critical point as the thermal entropy.
Soon after,  it is found that the entanglement entropy  can also probe the  Van der Waals-like phase transition in the $P-V$ plane \cite{Caceres}.
In \cite{Nguyen}, Nguyen has investigated exclusively the equal area law of holographic entanglement entropy and found that  the equal area law  holds regardless of the size of the entangling region. Very recently \cite{zeng2016} investigated entanglement entropy for a  quantum system with infinite volume, their result showed that   the  entanglement entropy also exhibits the  same Van der Waals-like phase transition as the thermal entropy. They also checked the equal area law and obtained the critical exponent of the  heat capacity near the critical point.

 In this paper, we will further investigate whether one can probe the phase structure by two point correlation function and Wilson loop besides the entanglement entropy.   We intend to explore whether they exhibit the similar Van der Waals-like phase transition as the entanglement entropy and thermal entropy. In addition, we also want to check whether these non-local observables can probe the  Hawking-Page phase transition between the AdS black hole and thermal gas so that  we can get a complete picture about the phase transition of the black holes in the framework of holography.

This paper is organized as follows. In the next section, we will discuss the thermal entropy phase transition of a 5-dimensional  Reissner-Nordstr\"{o}m-AdS  black hole in the $T-S$ plane in a fixed charge ensemble. Then in Section~\ref{Nonlocal_observables}, we will probe these phase transitions by  geodesic length, Wilson loop, and   holographic entanglement entropy individually. In each subsection, the  equal area law is checked and the critical exponent of the heat capacity is obtained for different sizes of the boundary region.
The last section is devoted to discussions and conclusions.

\section{Thermodynamic phase transition of the 5-dimensional   Reissner-Nordstr\"{o}m-AdS  black hole}

\label{quintessence_Vaidya_AdS}
Starting from the action
\begin{equation}
\label{eq1}
S =\frac{1}{16\pi G}\int d^{n+1}x \sqrt{-g} [ R-\frac{1}{4}F_{\mu\nu}F^{\mu\nu}+\frac{n(n-1)}{l^2}],
\end{equation}
where  $F_{\mu\nu}=\partial_{\mu} A_{\nu}-\partial_{\nu} A_{\mu}$, and $l$ is the AdS radius. We shall focus on the case of $n=4$, in which the charged Reissner-Nordstr\"{o}m-AdS  black hole can be written as \cite{Chamblin}
\begin{equation}
 ds^{2}=-f(r)dt^{2}+f^{-1}(r)dr^{2}+r^{2}[d\phi^2+\sin^2\phi (d\theta^2+\sin^2\theta d\psi^2)],\label{metric1}
\end{equation}
 where $\phi\in(0,\pi), \theta\in(0,\pi), \psi\in(0,2\pi)$,  are  hyperspherical coordinates for the 3-sphere, and  %
\begin{equation}
 f(r)=1+\frac{r^2}{l^2}-\frac{8 M}{3 \pi r^2}+\frac{4Q^2}{3 \pi^2 r^4}\label{metric},
\end{equation}%
with $M$ and $Q$ the mass and charge of the black hole. Whence we can get the Hawking temperature of this space time as
\begin{equation}
T=\frac{f^{\prime}(r)}{4\pi}\mid_{r_+}=\frac{3 \pi ^2 r_+^6-8 l^2 \left(Q^2-M \pi  r_+^2\right)}{6 l^2 \pi ^3 r_+^5}.\label{temperature}
 \end{equation}
 In addition,  it follows from the Bekenstein-Hawking formula that the entropy of the black hole is given by
\begin{equation}
S=\frac{\pi^2 r_+^3}{2},\label{entropy}
 \end{equation}
where $r_+$ is the outer event horizon of the black hole, namely the largest root of the equation $f(r_+)=0$. With this, the mass of the back hole can thus be expressed as the function of the event horizon
\begin{equation}
M=\frac{4 l^2 Q^2+3 l^2 \pi ^2 r_+^4+3 \pi ^2 r_+^6}{8 l^2 \pi  r_+^2}.\label{eentropy}
 \end{equation}
Substituting (\ref{entropy}) and (\ref{eentropy}) into (\ref{temperature}), we can get the relation between the temperature $T$ and entropy $S$ of the 5-dimensional   Reissner-Nordstr\"{o}m-AdS  black hole
\begin{equation}
T=\frac{12 S^2+l^2 \left(-2 \pi ^2 Q^2+3 2^{1/3} \pi ^{4/3} S^{4/3}\right)}{6 2^{2/3} l^2 \pi ^{5/3} S^{5/3}}.\label{temperatures}
 \end{equation}
In addition, with the relation  $F=M-TS$, the  Helmholtz free energy can be expressed as
\begin{equation}
F=\frac{5 Q^2}{6 \pi  r_+^2}-\frac{1}{8} \pi  r_+^2 \left(-1+r_+^2\right).\label{freeenergy}
 \end{equation}
Note that this formula for our free energy has implicitly chosen the pure AdS as the reference spacetime because the free energy vanishes for pure AdS by this formula. Now let us review the relevant phase transitions in the fixed charge ensemble by  (\ref{temperatures}) and (\ref{freeenergy}) in the  $T-S$ plane.

To achieve this, we should first find the critical charge by the following equations
\begin{equation}
(\frac{\partial T}{\partial S})_Q=(\frac{\partial^2 T}{\partial S^2})_Q=0. \label{heat}
 \end{equation}
Inserting   (\ref{temperatures}) into (\ref{heat}),
we can get the values for the critical charge and critical entropy
\begin{equation}
Q_c=\frac{l^2 \pi }{6 \sqrt{5}},
 \end{equation}
\begin{equation}
S_c=\frac{l^3 \pi ^2}{6 \sqrt{3}}. \label{criticalentropy}
 \end{equation}
 Substituting these critical values into  (\ref{temperatures}), we can get the critical temperature
\begin{equation}
T_c=\frac{\sqrt{3} (3 +5 )}{10 l \pi }\label{ct}.
 \end{equation}

\begin{figure}
\centering
\subfigure[$Q=0$]{
\includegraphics[scale=0.75]{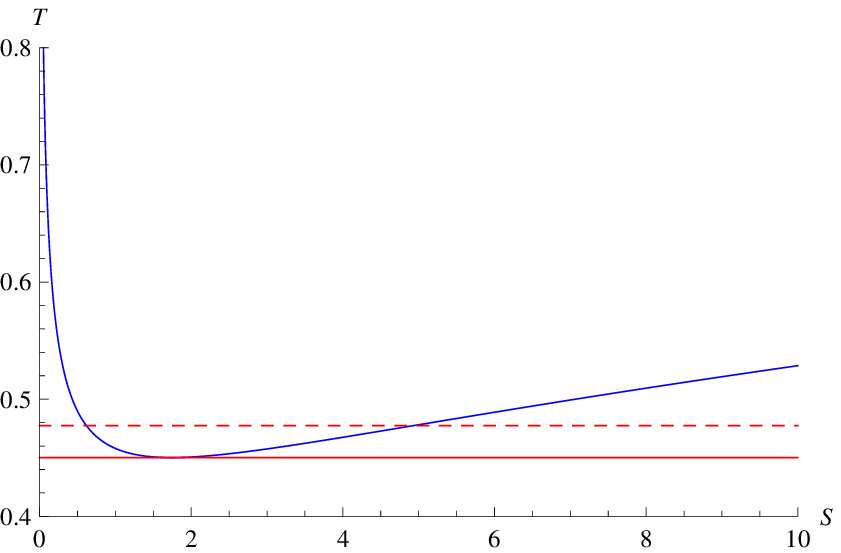}  }
\subfigure[$    Q=\frac{\pi -1}{6 \sqrt{5}}$]{
\includegraphics[scale=0.75]{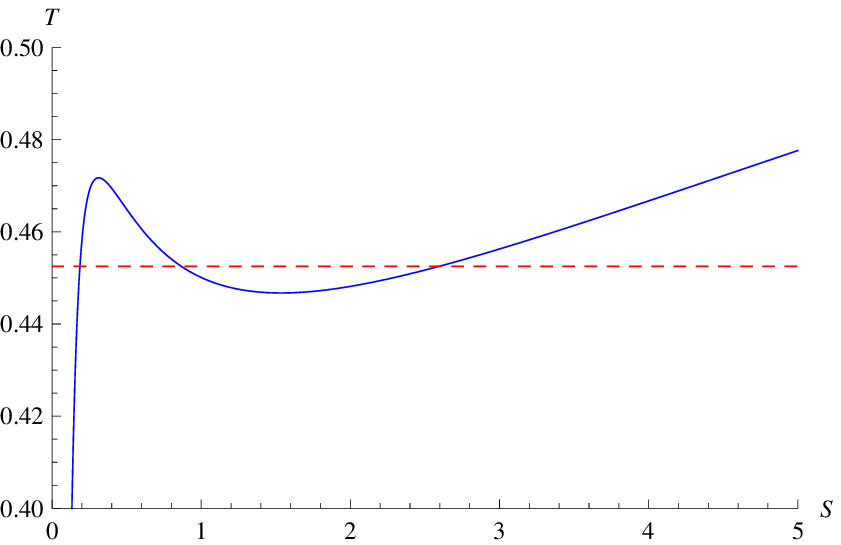}
}
\subfigure[$  Q=\frac{\pi}{6 \sqrt{5}}$]{
\includegraphics[scale=0.75]{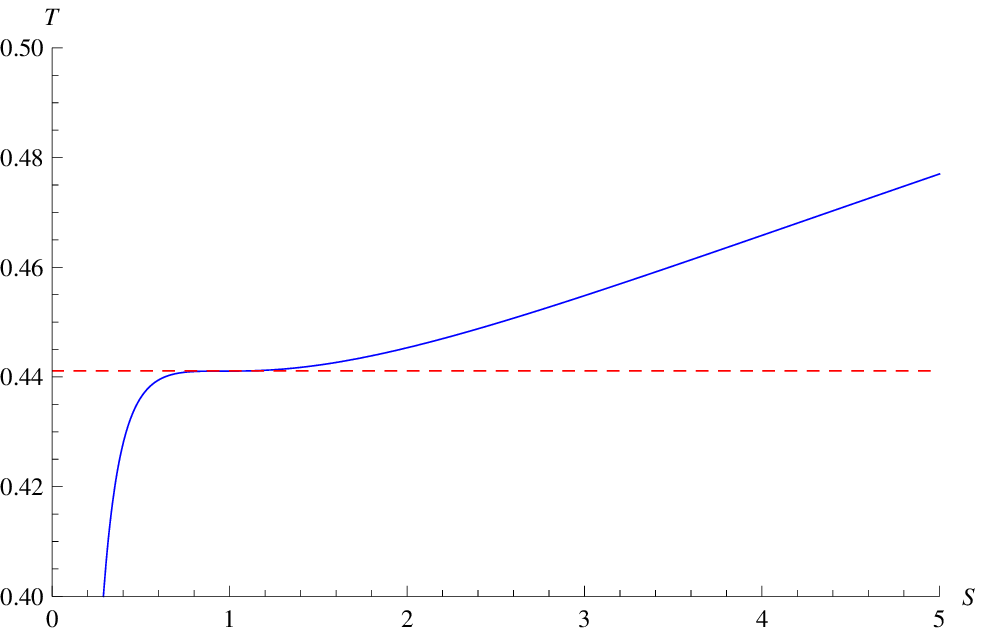}  }
\subfigure[$  Q=\frac{\pi+1}{6 \sqrt{5}}$]{
\includegraphics[scale=0.75]{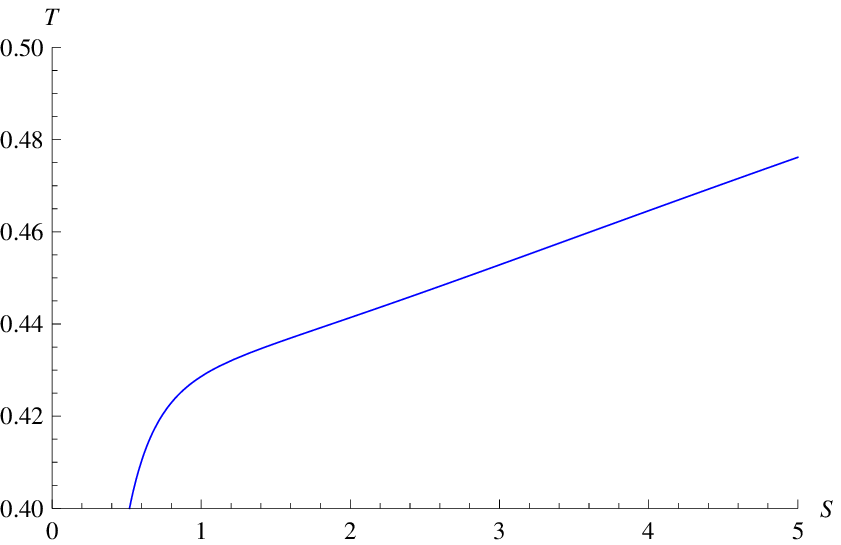}
}
 \caption{\small  Relation between the entropy and temperature for different charges in the fixed charge ensemble. The red solid line corresponds to the minimum temperature of space time and  the red dashed  lines  in (a), (b), (c) correspond to the locations of Hawking-Page phase transition, first order phase transition, and second order phase transition individually.} \label{fig1}
\end{figure}
\begin{figure}
\centering
\subfigure[$Q=0$]{
\includegraphics[scale=0.55]{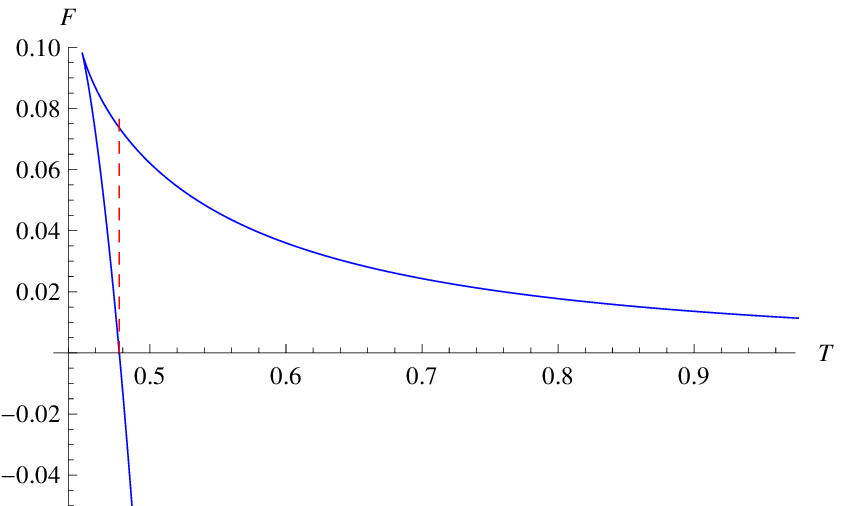}  }
\subfigure[$ Q=\frac{\pi -1}{6 \sqrt{5}}$]{
\includegraphics[scale=0.55]{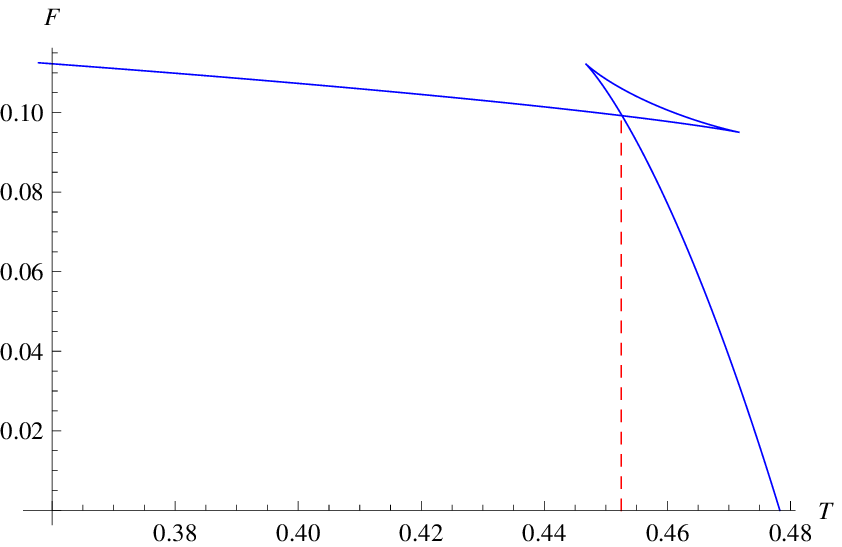}
}
\subfigure[$ Q=\frac{\pi}{6 \sqrt{5}}$]{
\includegraphics[scale=0.55]{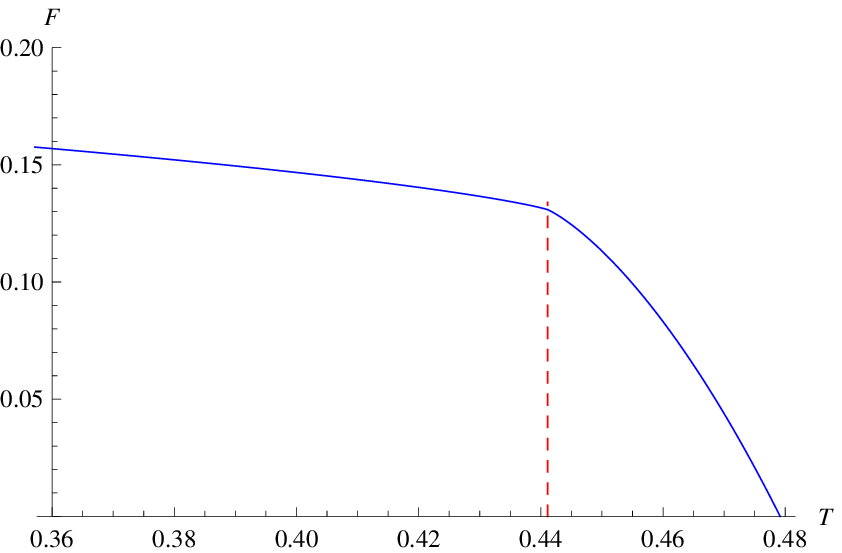}  }
 \caption{\small Relation between the free energy and temperature for different charges. The horizontal coordinates of the red dashed  lines correspond to the temperatures of the Hawking-Page phase transition,  first order phase transition, and second order phase transition.} \label{fig2}
\end{figure}

We plot the isocharge curves  for different charges in Figure \ref{fig1}. For the case $Q=0$,  There is a minimum temperature $T_0=\frac{\sqrt{2}}{ \pi }$ \cite{Banerjee}, which is indicated by the red solid line in $(a)$. When the temperature is lower than $T_0$,  we have only a thermal AdS. When the temperature is higher than $T_0$, there are two additional black hole branches. The small branch is unstable while the large branch is stable. This can be justified by check the corresponding heat capacities, which is related to their slopes. The Hawking-Page phase transition occurs at the temperature given by   $T_1=\frac{3}{2 \pi }$ \cite{Banerjee}, which is higher than $T_0$ and indicated by the red dashed line. This can be observed by  the $F-T$ relation in $(a)$ of Figure \ref{fig2},  where $T_0$ is the horizontal  coordinate of the cusp and  $T_1$ is the horizontal coordinate for the intersection of the stable branch and the horizontal axis. Obviously, when the temperature is lower than $T_1$, the thermal AdS is the most stable state. While when the temperature is higher than $T_1$, the most stable state is taken over by the large black hole branch.

 For the case $Q\neq 0$, the phase structure is
similar to that of the Van der Waals phase transition.
 That is, for a small charge, there is an unstable black hole interpolating between the stable small hole and stable large
 hole.  The small stable hole will jump to the large stable hole at the critical temperature temperature $T_{\star}$, which is labeled by the red dashed line in (b) of Figure \ref{fig1}.
  As the charge increases to the critical charge, the small hole and the large hole merge into one and squeeze  out the unstable phase. So there is an inflection point in $(c)$ of Figure \ref{fig1}. The heat capacity is divergent in this case, the phase transition is  therefore second order.
  As the charge exceeds the critical charge, we simply have one stable black hole at each temperature, which can be justified by the slope of the curve in $(d)$ of Figure  \ref{fig1}. The Van der Waals-like  phase transition  can also be observed from  the $F-T$ relation. From $(b)$ of  Figure \ref{fig2}, we see a swallowtail structure, which corresponds to the unstable phase in $(b)$ of Figure \ref{fig1}.  The critical temperature $T_{\star}=0.4526$ for the  phase transition is apparently read off by the horizontal coordinate of the junction between the small black hole and the large black hole. As the temperature  is lower than the critical temperature  $T_{\star}$, the free energy of the small black hole is lowest, so the small hole is stable. As the temperature is higher than $T_{\star}$, the free energy of the large black hole is lowest, so the large hole dominates thereafter. The non-smoothness of the junction indicates that the phase transition is first order.  When the charge is arriving at the critical charge $Q_c$, the swallowtail structure in
$(b)$ of  Figure \ref{fig2} shrinks into a point as is shown in $(c)$ of  Figure \ref{fig2}. The horizontal coordinate of the inflection point corresponds to the critical temperature $T_c$ of the second order phase transition, which is consistent with the analytical result in (\ref{ct}).

For the first order phase transition in $(b)$ of Figure \ref{fig1}, we would like to check whether Maxwell's equal area law holds with the following formula
\begin{equation}
A_1\equiv\int_{S_1}^{S_3}T(S,Q)dS=T_{\star}(S_3-S_1)\equiv A_3, \label{euqalarea}
 \end{equation}
in which $T(S,Q)$ is defined in  (\ref{temperatures}), $S_1$ and  $S_3$ are the smallest and largest roots of the equation $T(S,Q)=T_{\star}$.
After a simple calculation, we find   $S_1=0.186987$ and $S_3=2.60575$. With these values, we find $A_1$ and $A_3$ in (\ref{euqalarea}) equal 1.09481 and 1.09482 respectively. So the equal area law in the $T-S$ plane holds within our numerical accuracy.

For the second order phase transition in $(c)$ of Figure \ref{fig1}, we are interested in the critical exponent associated with the heat capacity
 \begin{equation}
C_{Q}=T\frac{\partial S}{\partial T}\mid_Q \label{capacity}.
 \end{equation}
 Near the critical point, writing the entropy as $S=S_c+\delta$  and expanding the temperature in terms of small $\delta$, we find
\begin{eqnarray}
T-T_c=\frac{\left(220 l^2 \pi ^2 Q^2-21 2^{1/3} l^2 \pi ^{4/3} S^{4/3}+30 S^2\right) }{243 2^{2/3} l^2 \pi ^{5/3} S^{14/3}}(S-S_c)^3\label{c1},
 \end{eqnarray}
in which  we have used (\ref{heat}).
In this case, (\ref{capacity}) further implies  $C_Q\sim(T-T_c)^{-2/3}$, namely the critical exponent is $-2/3$, which is the same as the one from the mean field theory. In addition, taking  logarithm to (\ref{c1}), we have a linear relation
\begin{eqnarray}
\log\mid T-T_c\mid =3 \log\mid S- S_c\mid +constant  \label{c2},
 \end{eqnarray}
 with $3$ the slope. In what follows,  we will use this logarithm to check the critical exponent for the analogous heat capacities in the framework   of holography.

It it noteworthy that by holography the whole phase structure described above is not only for the bulk black hole but also for the dual boundary system, where the thermal entropy is simply given by the black hole entropy, and so on so forth.

\section{Phase transition in the framework of holography}
\label{Nonlocal_observables}
In this section we shall investigate the phase structures of some non-local observables such as two point  correlation function, Wilson loop, and entanglement entropy in the dual fled theory by holography to see whether they have the same phase structure as the thermal entropy.

\subsection{Phase transition of two point correlation function}

 According to the AdS/CFT correspondence,  if the conformal dimension $\Delta$ of scalar operator $\cal{O}$ of dual field theory
is large enough, the equal time two point correlation function  can
be holographically approximated as \cite{Balasubramanian61}
\begin{equation}
\langle {\cal{O}} (t_0,x_i) {\cal{O}}(t_0, x_j)\rangle  \approx
e^{-\Delta {L}} ,\label{llll}
\end{equation}
where
$L$ is the  length of the bulk geodesic between the points $(t_0,
x_i)$ and $(t_0, x_j)$ on the AdS boundary.
 Taking into account the spherical symmetry of the 5-dimensional Reissner-Nordstr\"{o}m-AdS  black hole,
  we can simply choose $(\phi=\frac{\pi}{2}, \theta=\theta_0, \psi=0)$ and $(\phi=\frac{\pi}{2},\theta=\theta_0,\psi=\pi)$ as the two boundary points. Then with  $\theta$ to
    parameterize the trajectory, the proper length  is given by
\begin{eqnarray}
L=2\int_0 ^{\theta_0}\mathcal{L}(r(\theta),\theta) d\theta,~~\mathcal{L}=\sqrt{\frac{\dot{r}^2}{f(r)}+r^2},
 \end{eqnarray}
in which $\dot{r}=dr/ d\theta$.
Imagining $\theta$ as time, and treating $\mathcal{L}$ as the Lagrangian, one can
get the equation of motion for $r(\theta)$ by making use of the Euler-Lagrange equation, that is
\begin{eqnarray}
0=\dot{r}^2 f'(r)-2 f(r) \ddot{r}+2 r f(r)^2,
\end{eqnarray}
which can be solved by imposing the following  boundary conditions
\begin{eqnarray}
\dot{r}(0)=0, r(0)= r_0.\label{bon}
\end{eqnarray}
To explore whether the size of the boundary region affects the later phase structure,  we here choose  $\theta_0=0.14, 0.2$  as two examples. Note that for a fixed $\theta_0$, the geodesic length  is divergent, so it should be regularized by subtracting off the geodesic length in pure AdS with the same  boundary region, denoted by $L_0$. To achieve this, we are required to set a UV cutoff for each case, which is chosen to be $ r(0.139)$ and $ r(0.199)$, respectively for our two examples.  In this paper, we obtain $L_0$  also by numerics though there
 is an analytical result for $r(\theta_0)$ for pure AdS in Einstein gravity.
We label the regularized  geodesic length as $\delta L\equiv L-L_0$.
\begin{figure}
\centering
\subfigure[$Q=0$]{
\includegraphics[scale=0.75]{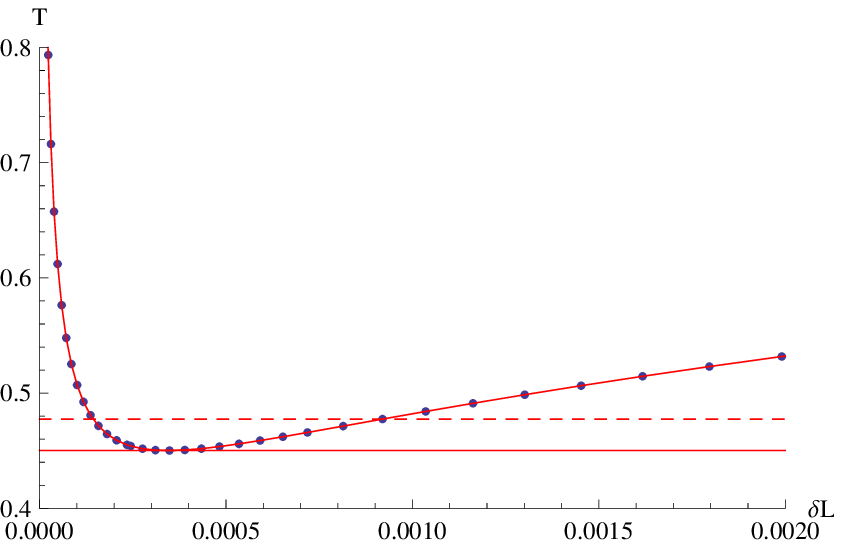}  }
\subfigure[$Q=\frac{\pi -1}{6 \sqrt{5}}$]{
\includegraphics[scale=0.75]{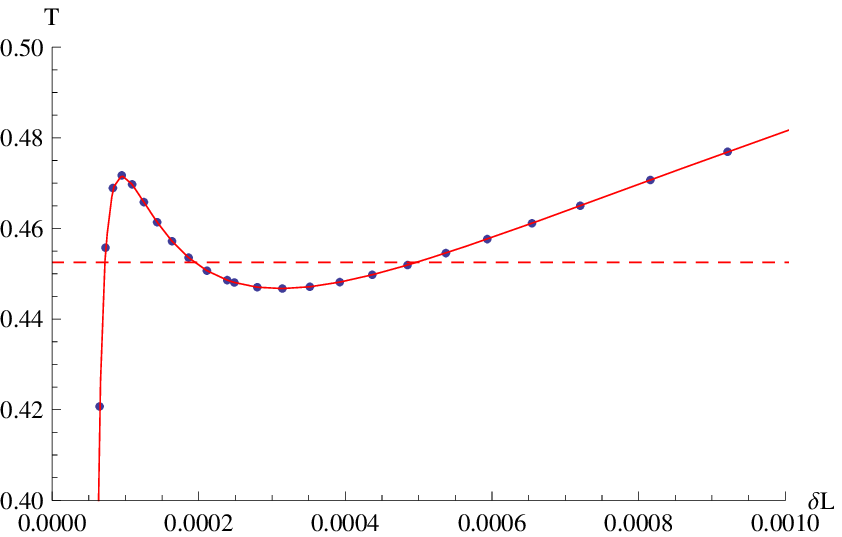}
}
\subfigure[$Q=\frac{\pi}{6 \sqrt{5}}$]{
\includegraphics[scale=0.75]{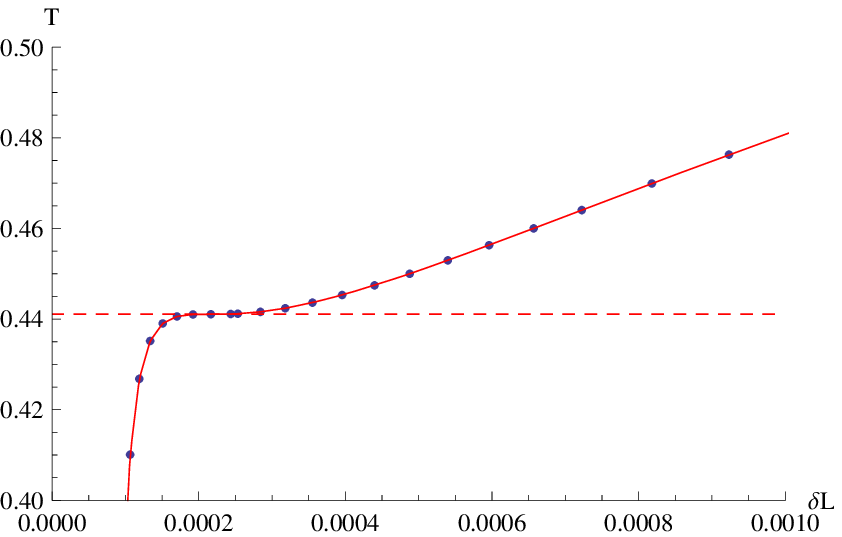}
}
\subfigure[$Q=\frac{\pi +1}{6 \sqrt{5}}$]{
\includegraphics[scale=0.75]{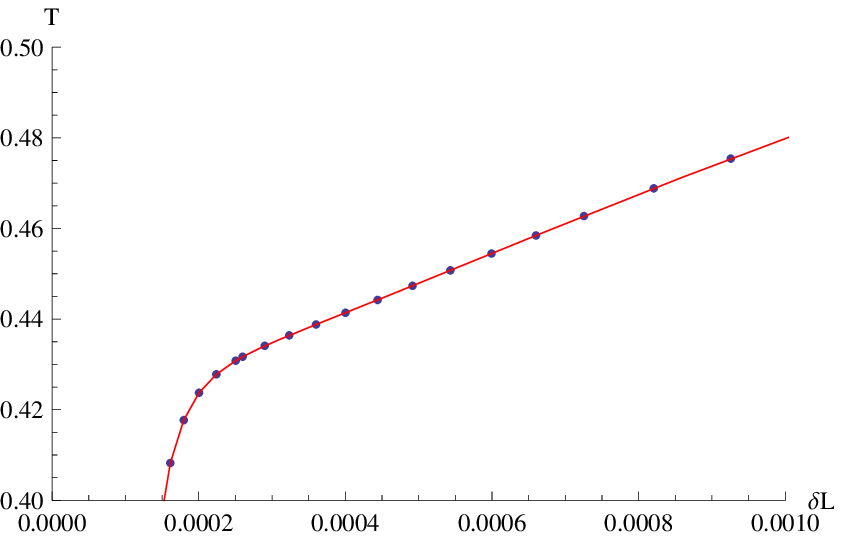}
}
\caption{\small Relation between the geodesic length and temperature  in the fixed charge ensemble for differen charges at $\theta_0=0.14$. The red solid line corresponds to the location of the minimum temperature $T_0$,  the  dashed  lines  in (a), (b), (c) correspond individually to the locations of Hawking-Page phase transition $T_1$, first order phase transition $T_{\star}$, and second order phase transition $T_c$.} \label{fig3}
\end{figure}

\begin{figure}
\centering
\subfigure[$Q=0$]{
\includegraphics[scale=0.75]{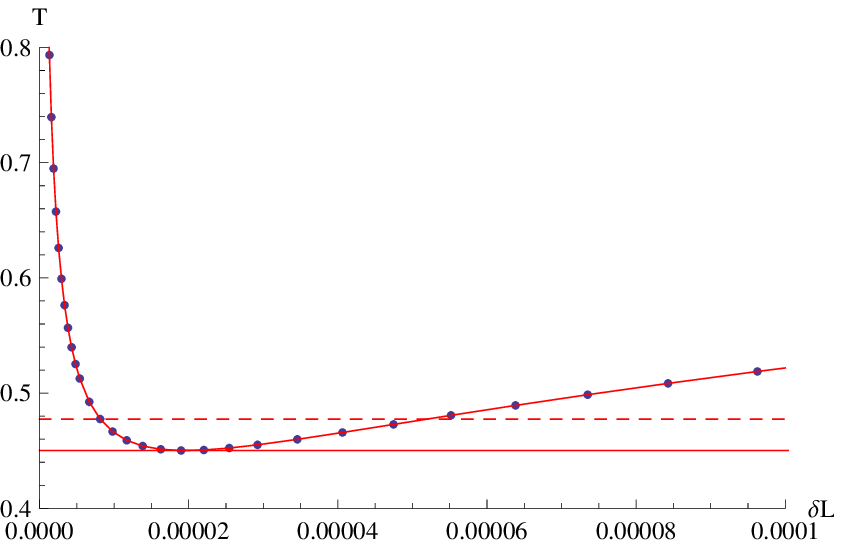}  }
\subfigure[$Q=\frac{\pi -1}{6 \sqrt{5}}$]{
\includegraphics[scale=0.75]{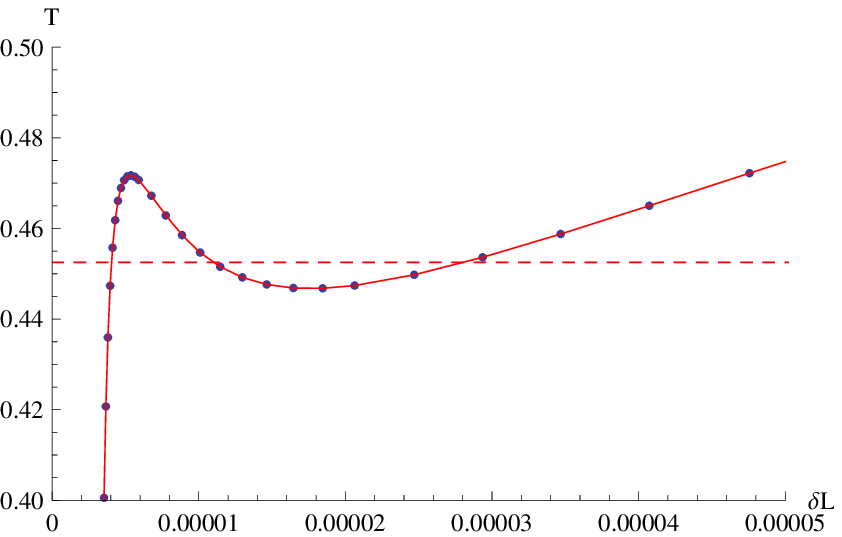}
}
\subfigure[$Q=\frac{\pi }{6 \sqrt{5}}$]{
\includegraphics[scale=0.75]{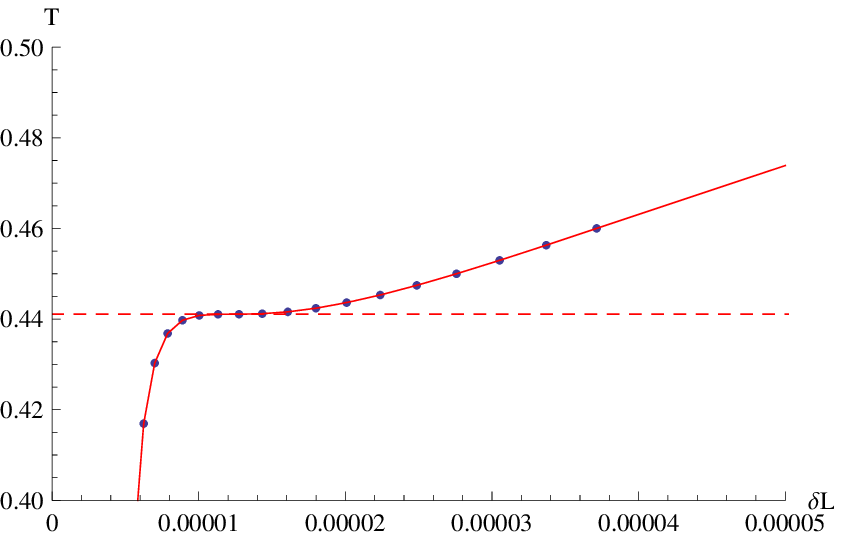}
}
\subfigure[$Q=\frac{\pi +1}{6 \sqrt{5}}$]{
\includegraphics[scale=0.75]{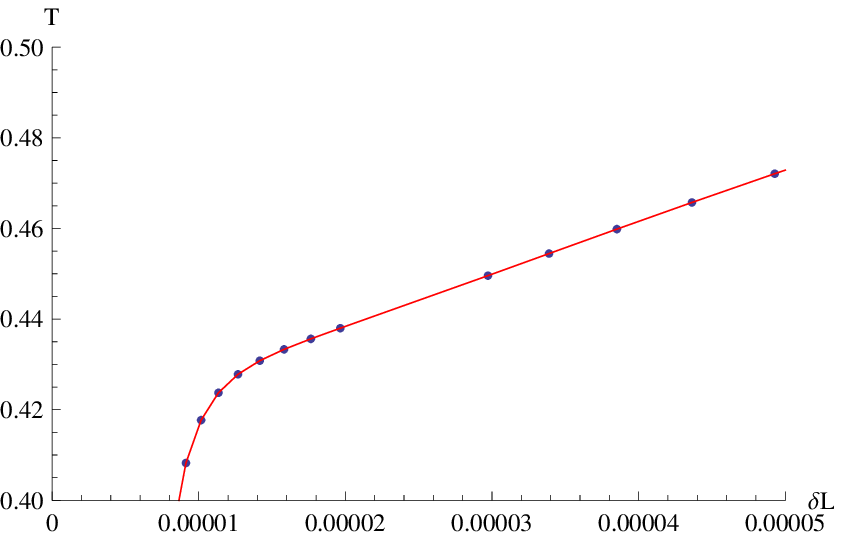}
}
\caption{\small Relation between the geodesic length and temperature  in the fixed charge ensemble for differen charges at $\theta_0=0.2$.  The red solid line corresponds to the location of the minimum temperature $T_0$,  the  dashed lines  in (a), (b), (c) correspond individually to the locations of Hawking-Page phase transition $T_1$, first order phase transition $T_{\star}$, and second order phase transition $T_c$.} \label{fig4}
\end{figure}

We plot the relation between  $T$ and $\delta L$ for different  $\theta_0$  in Figure \ref{fig3} and  Figure \ref{fig4}.
 As shown in Figure \ref{fig3}  and Figure \ref{fig4},  $\delta L$ demonstrates a similar phase structure as the thermal entropy. Moreover,
 we find that  the minimum temperature $T_0$ as well as Hawking-Page phase transition temperature  $T_1$ in $(a)$,  the first order phase transition temperature  $T_{\star}$ in $(b)$, and second order phase transition temperature  $T_c$ in $(c)$ are exactly the same as those in $T-S$ plane, which justifies our notation. To be more specific, it is easy to check $T_0$ by locating the position of local minimum. But in order to confirm $T_{\star}$ and $T_c$, we are required to examine the equal area law for the first order phase transition and obtain $-2/3$ as the critical exponent for the second order phase transition, which are documented as follows.

  In the  $\delta L-T$ plane, we define the  equal area law  as
\begin{eqnarray}
A_1\equiv\int_{\delta L_1}^{\delta L_3} T(\delta L) d\delta L= T_{\star} (\delta L_3-\delta L_1)\equiv A_3,\label{arealaw}
 \end{eqnarray}
in which  $T(\delta L)$  is an Interpolating Function obtained from the numeric result, and $\delta L_1$,  $\delta L_3$ are the smallest and largest roots of the equation $T(\delta L)=T_{\star}$.
For the case $\theta_0=0.14$, we find   $\delta L_1=0.0000556147$,  $\delta L_3=0.000194497$. Substituting these values into  (\ref{arealaw}), we find  $A_1=0.0000628401$,  $A_3=0.0000628583$. For the case $\theta_0=0.2$, after simple calculation, we find
 $A_1=0.0000108924$,  $A_3=0.0000108875$.
It is obvious that for different $\theta_0$, $A_1 $ and $A_3$ are equal within our numeric accuracy. Thus the equal area law also holds  in the $\delta L-T$ plane.

\begin{figure}
\centering
\subfigure[$\theta_0=0.14$]{
\includegraphics[scale=0.75]{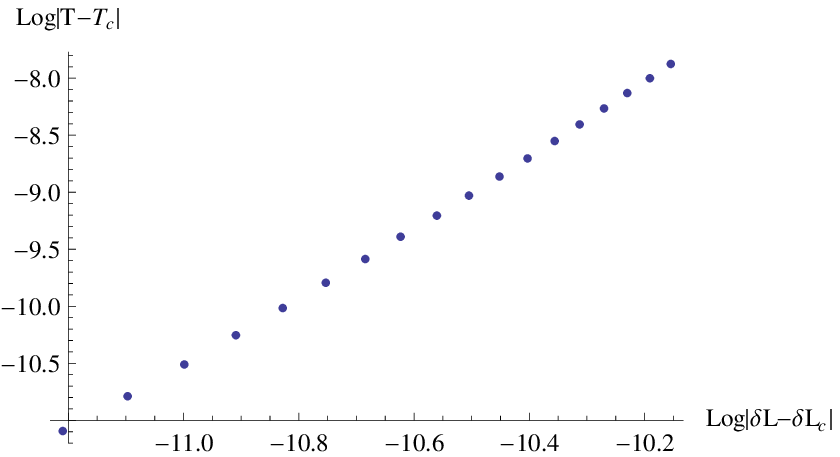}
 }
 \subfigure[$\theta_0=0.2$]{
\includegraphics[scale=0.75]{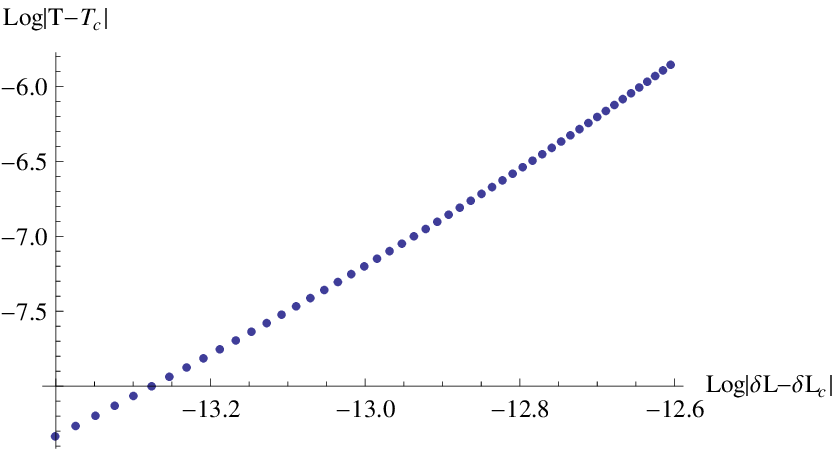}
 }
 \caption{\small Relation between   $\log\mid T-T_c\mid$ and $\log\mid\delta L-\delta L_c\mid $ near the critical point of second order phase transition for different $\theta_0$.} \label{fig5}
\end{figure}

In addition, in order to investigate the critical exponent for the analogous heat capacity of the geodesic length. we are interested in the logarithm of the quantities $T-T_c$, $\delta L-\delta L_c$, in which $T_c$ is the critical temperature defined in  (\ref{ct}), and
 $L_c$ is  obtained numerically by the equation $T(\delta L)=T_c$.
 We plot the relation between $ \log\mid T -T_c\mid$ and $\log\mid\delta L-\delta L_c\mid  $ for different $\theta_0$  in Figure \ref{fig5}, where
 these
straight lines can be fitted as
\begin{equation}
\log\mid T-T_c\mid=\begin{cases}
23.2318 + 3.06832  \log\mid\delta L-\delta L_c\mid,&  $for$ ~\theta_0=0.14,\\
31.9841 + 3.00077  \log\mid\delta L-\delta L_c\mid, &~$for$~\theta_0=0.2.\\
\end{cases}
\end{equation}
It is obvious that the slope is about 3, which indicates that the critical exponent is $-2/3$ for the analogous heat capacity and the phase transition is also second order at $T_c$ for the geodesic length.

\subsection{Phase transition of Wilson loop}

\begin{figure}
\centering
\subfigure[$Q=0$]{
\includegraphics[scale=0.75]{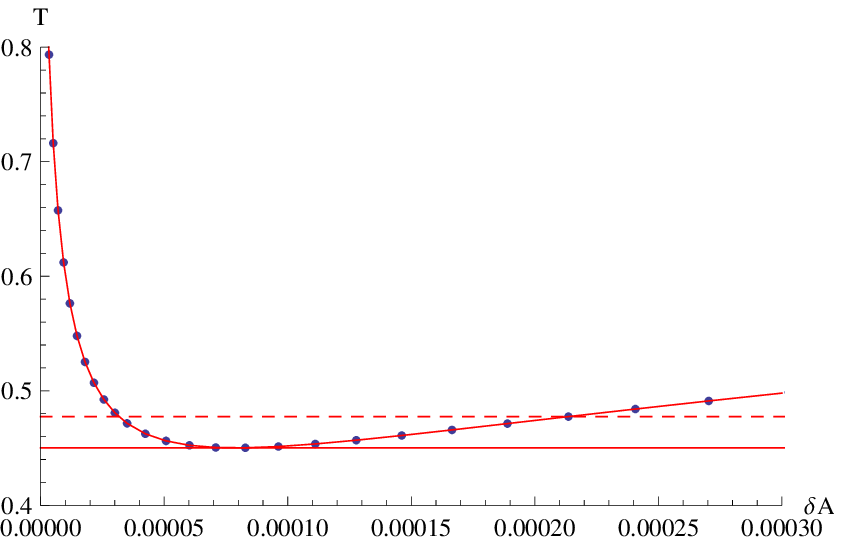}  }
\subfigure[$Q=\frac{\pi -1}{6 \sqrt{5}}$]{
\includegraphics[scale=0.75]{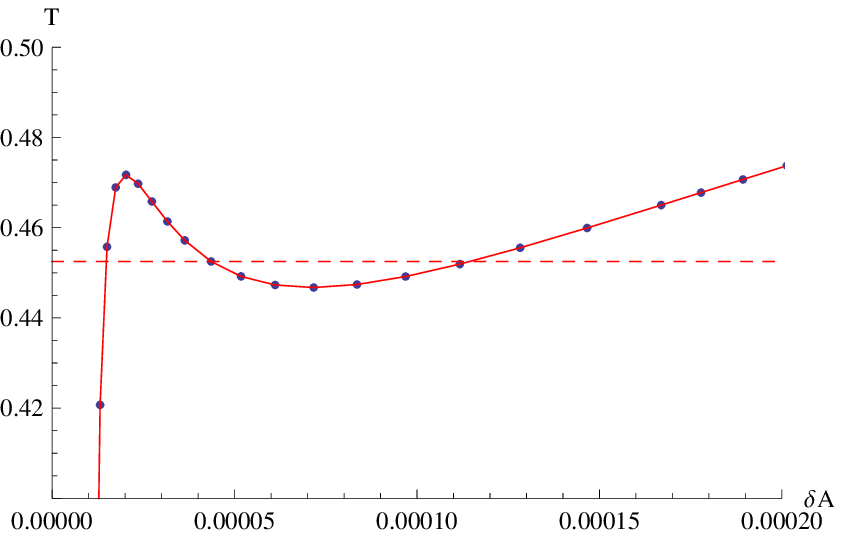}
}
\subfigure[$Q=\frac{\pi}{6 \sqrt{5}}$]{
\includegraphics[scale=0.75]{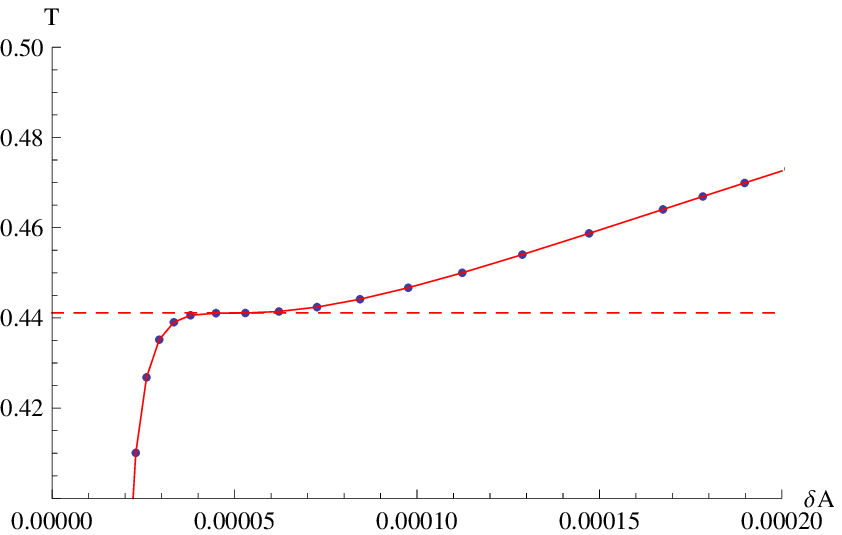}
}
\subfigure[$Q=\frac{\pi +1}{6 \sqrt{5}}$]{
\includegraphics[scale=0.75]{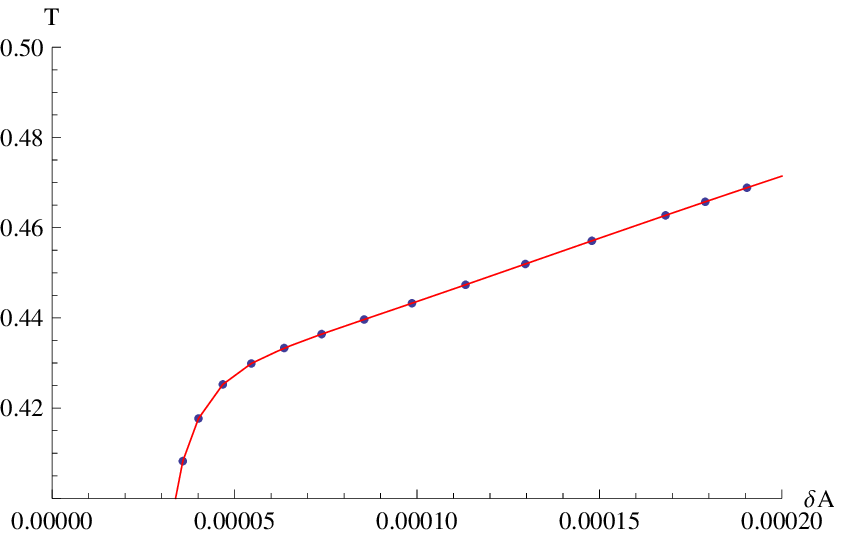}
}
\caption{\small Relation between the  minimal area surface and temperature  in the fixed charge ensemble for differen charges at $\theta_0=0.14$.  The red solid line corresponds to the location of the minimum temperature $T_0$,  the  dashed  lines  in (a), (b), (c) correspond individually to the locations of Hawking-Page phase transition $T_1$, first order phase transition $T_{\star}$, and second order phase transition $T_c$.} \label{fig6}
\end{figure}

\begin{figure}
\centering
\subfigure[$Q=0$]{
\includegraphics[scale=0.75]{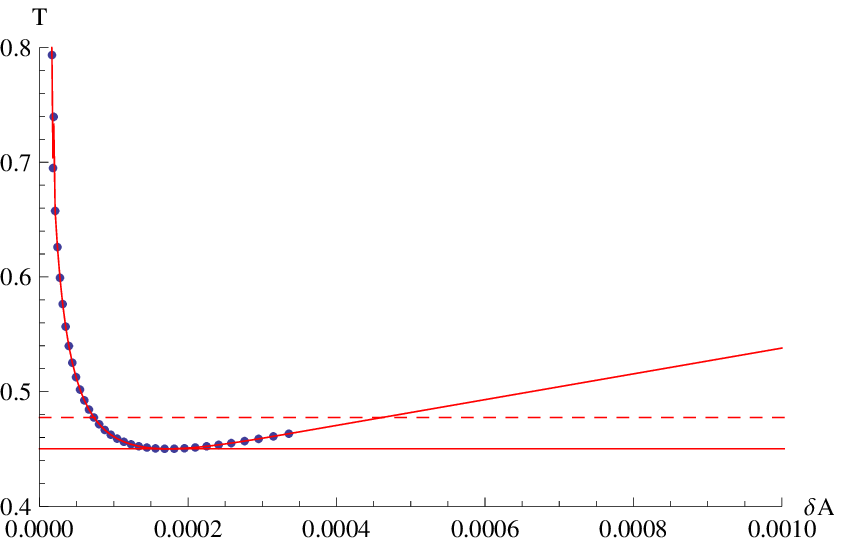}  }
\subfigure[$Q=\frac{\pi-1}{6 \sqrt{5}}$]{
\includegraphics[scale=0.75]{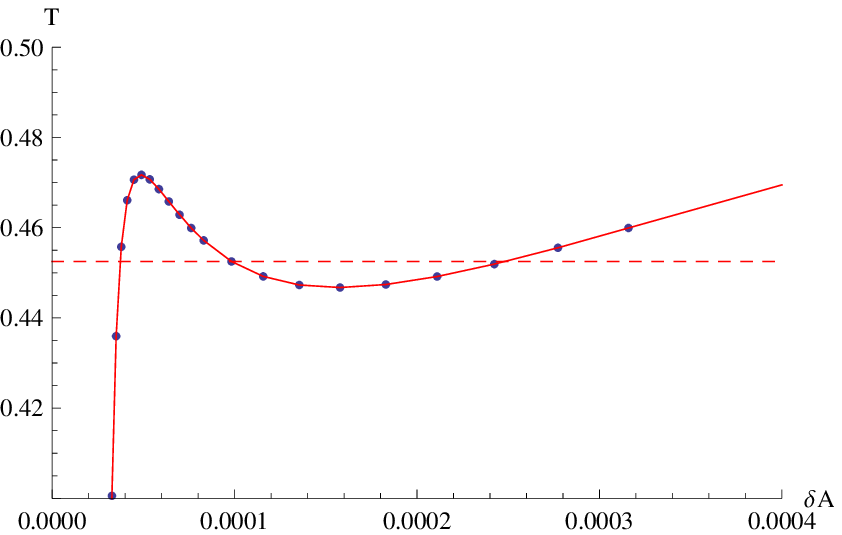}
}
\subfigure[$Q=\frac{\pi }{6 \sqrt{5}}$]{
\includegraphics[scale=0.75]{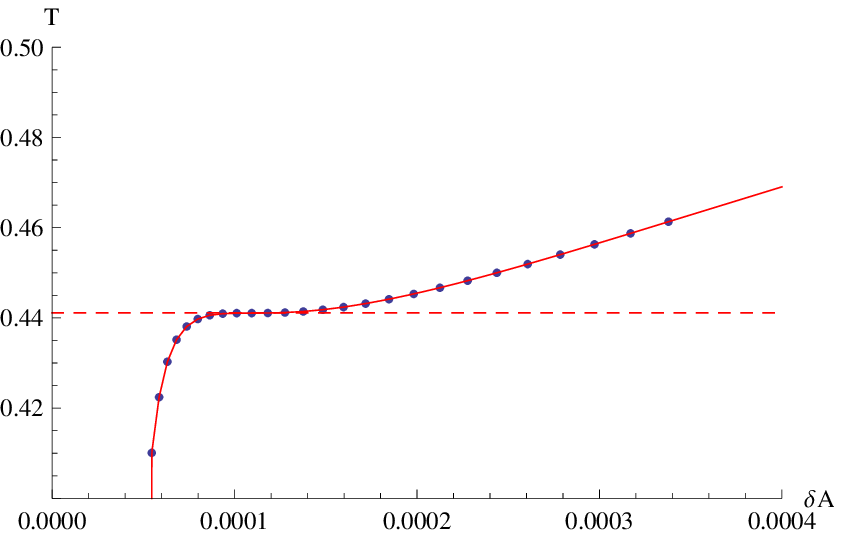}
}
\subfigure[$Q=\frac{\pi +1}{6 \sqrt{5}}$]{
\includegraphics[scale=0.75]{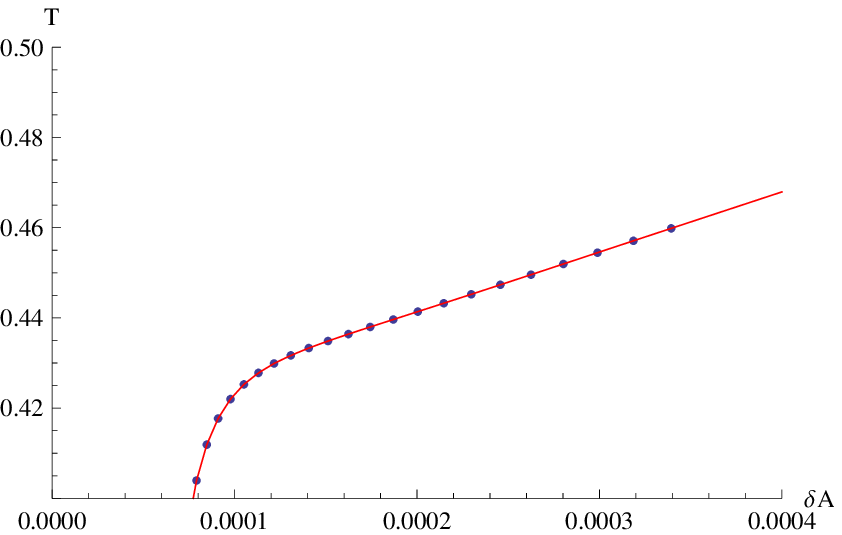}
}
\caption{\small Relation between the  minimal area surface and temperature  in the fixed charge ensemble for differen charges at $\theta_0=0.2$.  The red solid line corresponds to the location of the minimum temperature $T_0$,  the  dashed  lines  in (a), (b), (c) correspond individually to the locations of Hawking-Page phase transition $T_1$, first order phase transition $T_{\star}$, and second order phase transition $T_c$.} \label{fig7}
\end{figure}

In this subsection, we are going to study the phase structure of the Wilson loop,  which in the bulk corresponds to
the minimal area surface by holography.
 Wilson loop operator is defined as a path ordered integral of gauge field over a closed contour,
 and its expectation value is approximated geometrically  by the AdS/CFT correspondence as \cite{ Maldacena80}
\begin{equation} \label{area}
\langle W(C)\rangle \approx e^{-\frac{A_\Sigma}{2\pi\alpha'}},
\end{equation}
where $C$ is the closed contour, $\Sigma$ is the minimal bulk surface ending on $C$ with $A$ its
minimal  area, and  $\alpha'$ is the Regge slope parameter.
Next we choose the line with $\phi=\frac{\pi}{2}$ and $\theta=\theta_0$ as our loop. Then we can employ  $(\theta, \psi)$  to parameterize the minimal area surface, which is invariant under the $\psi$-direction by our rotational symmetry.
 Thus the
corresponding minimal area surface   can be expressed as
\begin{eqnarray}
A=2\pi \int_0 ^{\theta_0} r \sin \theta \sqrt{\frac{\dot{r}^2}{f(r)}+r^2}   d\theta,
 \end{eqnarray}
in which $\dot{r}=dr/ d\theta$.
Making use of the Euler-Lagrange equation, one can get the equation of motion for $r(\theta)$. Then with the boundary conditions  $r^{\prime}(0)=0$, $r(0)= r_0$, we can further get the numeric result of  $r(\theta)$.
Similar to the case of geodesic length, we choose  $\theta_0=0.14, 0.2$  as two examples and  the corresponding UV cutoffs are set  to be $ r(0.139)$, $ r(0.199)$.
 We label the regularized minimal area surface as $\delta A\equiv A-A_0$, where  $A_0$ is the minimal area in pure AdS with the same  boundary region.
  We plot the relation between  $\delta A$ and $T$  for different  $\theta_0$ in Figure \ref{fig6} and  Figure \ref{fig7}. Comparing  Figure \ref{fig6} with  Figure \ref{fig7}, we find they are the same nearly besides the scale of the horizonal coordinate. In other words,    $\theta_0$ affects only the value but not the phase structure of minimal area surface in the $T-\delta A$ plane. The result tells us that the similar phase structure also shows up for the minimal surface area. Here we concentrate only on scrutinizing  the equal area law for the first order phase transition and  the critical exponent of the analogous heat capacity for the second order phase transition.

First, in the  $\delta A-T$ plane,  the  equal area law can be similarly defined  as
\begin{eqnarray}
A_1\equiv\int_{\delta A_1}^{\delta A_3} T(\delta A) d\delta A= T_{\star} (\delta A_3-\delta A_1)\equiv A_3,\label{1arealaw}
 \end{eqnarray}
in which  $T(\delta A)$  is an Interpolating Function obtained from our numeric result, and $\delta A_1$,  $\delta A_3$ are the smallest and largest roots of the equation $T(\delta A)=T_{\star}$, respectively. As the same as that of the geodesic length, for a fixed $\theta_0$, we first obtain $\delta A_1$,  $\delta A_3$, and then  substitute these values into  (\ref{1arealaw}) to produce $A_1$,  $A_3$.  The concrete values are listed in
Table \ref{tab1}.
\begin{table}
\begin{center}\begin{tabular}{c|c|c}
 %\MC{3}{c}{\text{caption}}\\[5pt]
 \hline
% & \multicolumn{3}{c||}{MGCDM}   & \multicolumn{3}{c}{$\Lambda$CDM}  \\ \hline
%                             &        MGCDM        &                  &             &      $\Lambda$CDM    &                   & \\ \hline
% \MC{3}{|c|c|}{\ZZ{-8pt}{15pt}\hfill\normalsize   \hfill  \hfill\normalsize MGCDM     \hfill\normalsize $\Lambda$CDM  }\\ \hline
% \ZZ{-6pt}{22pt}
 $\theta_0=0.14$ &                                          $\theta_0=0.2$         \\ \hline
                   $T_{\star}$ =0.4526              &    $T_{\star}$ =0.4526       \\    \hline
$\delta A_1$=0.0000434217$\mid$  $\delta A_3$=0.000114835&   $\delta A_1$=0.0000374703$\mid$  $\delta A_3$=0.000248923           \\\hline
$A_1$=0.0000320531$\mid$  $A_3$=0.0000323218&   $A_1$=0.0000957474$\mid$  $A_3$=0.0000957035
  \\ \hline
\end{tabular}
\end{center}
\caption{Check of the equal area law in the $T-\delta A$ plane for different $\theta_0$.}\label{tab1}
\end{table}
Obviously, for both $\theta_0$, $A_1$ and   $A_3$ are equal within the reasonable numeric accuracy. The equal area law thus holds in the $\delta A-T$ plane, which reinforces the fact that the minimal surface area has the same first order phase transition behavior as that of the thermal entropy.

\begin{figure}
\centering
\subfigure[$\theta_0=0.14$]{
\includegraphics[scale=0.75]{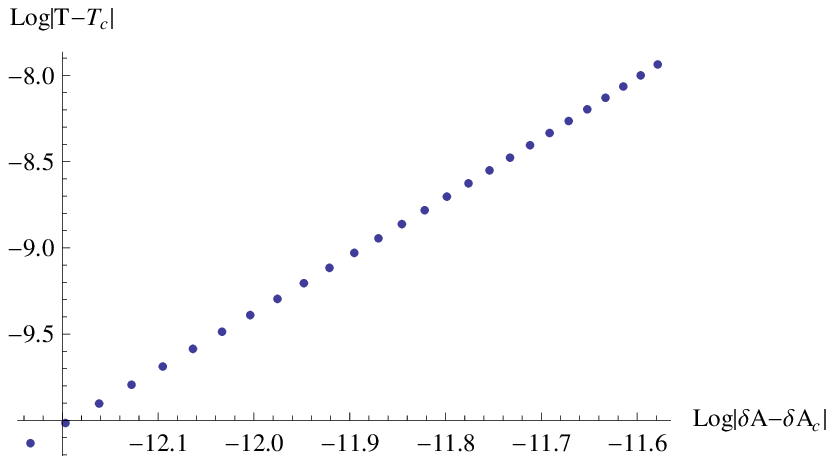}
 }
 \subfigure[$\theta_0=0.2$]{
\includegraphics[scale=0.75]{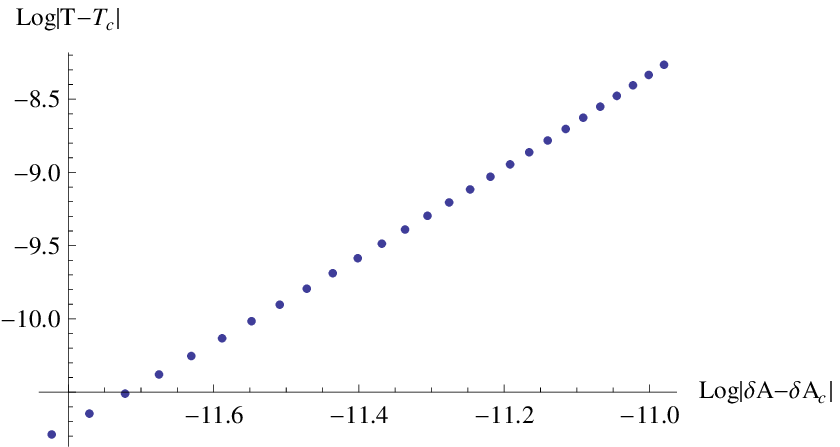}
 }
 \caption{\small Relation between  $\log\mid T-T_c\mid$ and $\log\mid\delta A-\delta A_c\mid $ near the critical point of second order phase transition for different $\theta_0$.} \label{fig8}
\end{figure}

Second, in order to check whether the minimal surface area also demonstrates the same second order phase transition as the thermal entropy, we would like to evaluate the critical exponent of the analogous heat capacity at the critical point in the $\delta A-T$ plane.
To this end, we plot the relations between $ \log\mid T -T_c\mid$ and $\log\mid\delta A-\delta A_c\mid  $  in Figure \ref{fig8}.
The numerical results  for these
curves can be fitted as
\begin{equation}
\log\mid T-T_c\mid=\begin{cases}
27.5226 + 3.04698  \log\mid\delta A-\delta A_c\mid,&  $for$ ~\theta_0=0.14,\\
24.692 + 3.00462  \log\mid\delta A-\delta A_c\mid, &~$for$~\theta_0=0.2.\\
\end{cases}
\end{equation}
With $3$ the slope, we can conclude that the minimal surface area also has the same second order phase transition as the thermal entropy.

\subsection{Phase transition of entanglement entropy}
Holographic entanglement entropy is another non-local observable, and  it has been used  extensively to probe the  superconductivity phase transition besides the thermalization process recently \cite{
Superconductors1, Superconductors2, Superconductors3, Superconductors4, Superconductors5, Superconductors7, Superconductors8,Ling}.  In this subsection, we intend to  employ it to probe the phase structure of a  5-dimensional Reissner-Nordstr\"{o}m-AdS  black hole.
 According to the formula in \cite{Ryu,Ryu1}, holographic entanglement entropy can be given by the area
$A_{\Sigma}$ of a minimal surface $\Sigma$ anchored on the boundary entangling surface $\partial \Sigma$, namely
\begin{equation}
S=\frac{A_{\Sigma}(t)}{4 G}. \label{eee}
\end{equation}
For simplicity, we choose  $\phi=\phi_0$ as our entangling surface and employ  $(\phi,\theta, \psi)$  to parameterize the minimal surface.
But with the symmetry of  (\ref{metric1}),
(\ref{eee}) can be rewritten as
\begin{eqnarray}
S=4 \pi \int_0 ^{\phi_0}r^2 \sin^2\phi\sqrt{\frac{\dot{r}^2}{f(r)}+r^2}d\phi
 \end{eqnarray}
 with $\dot{r}=dr/d\phi$.
Similarly, we can solve the equation of motion for $r(\phi)$ numerically, and eventually obtain the regularized entanglement entropy $\delta S$. We plot the relation between $\delta S$ and $T$ for $\phi_0=0.14, 0.2$  in
\begin{figure}
\centering
\subfigure[$Q=0$]{
\includegraphics[scale=0.75]{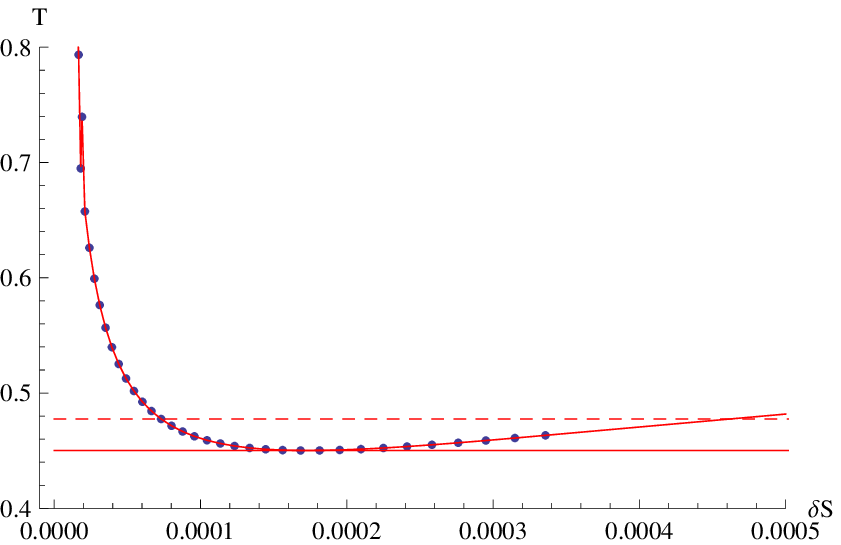}  }
\subfigure[$
Q=\frac{\pi -1}{6 \sqrt{5}}$]{
\includegraphics[scale=0.75]{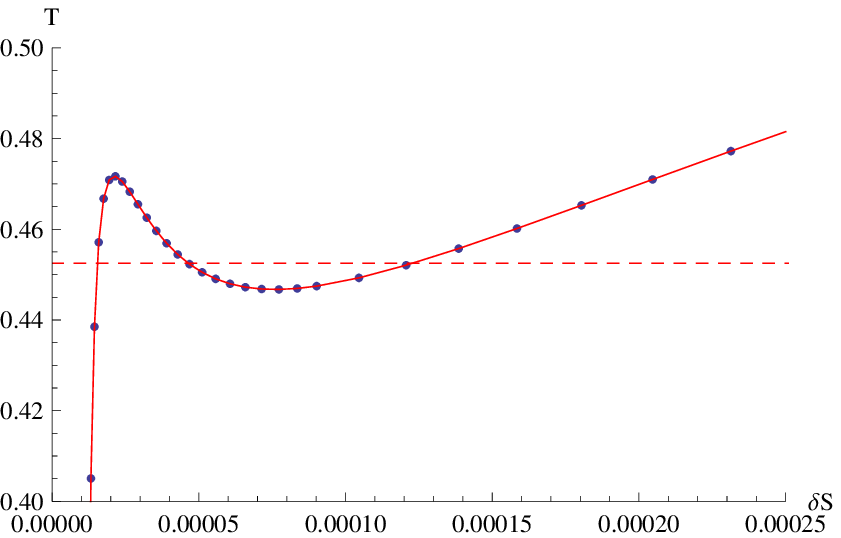}
}
\subfigure[$Q=\frac{\pi }{6 \sqrt{5}}$]{
\includegraphics[scale=0.75]{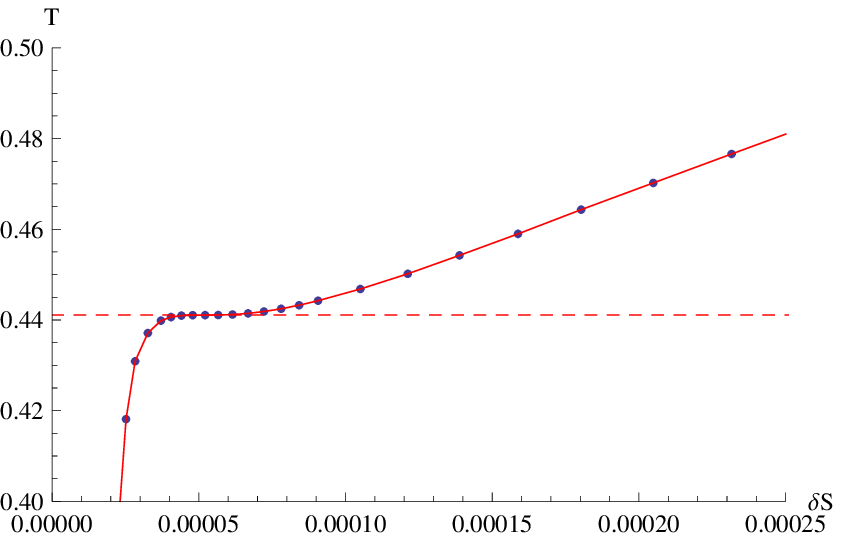}
}
\subfigure[$Q=\frac{\pi +1}{6 \sqrt{5}}$]{
\includegraphics[scale=0.75]{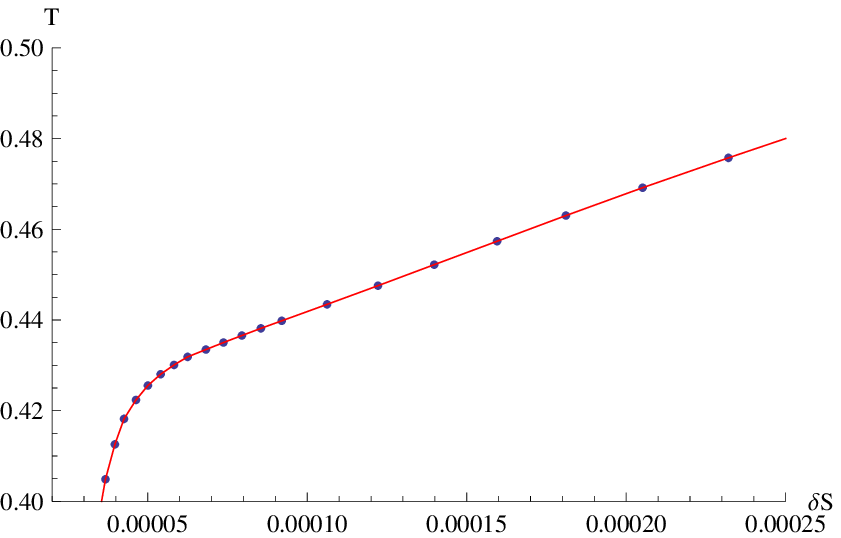}
}
\caption{\small Relation between the  entanglement entropy and temperature  in the fixed charge ensemble for differen charges at $\phi_0=0.14$.  The red solid line corresponds to the location of the minimum temperature $T_0$,  the  dashed  lines  in (a), (b), (c) correspond individually to the locations of Hawking-Page phase transition $T_1$, first order phase transition $T_{\star}$, and second order phase transition $T_c$.} \label{fig9}
\end{figure}
\begin{figure}
\centering
\subfigure[$Q=0$]{
\includegraphics[scale=0.75]{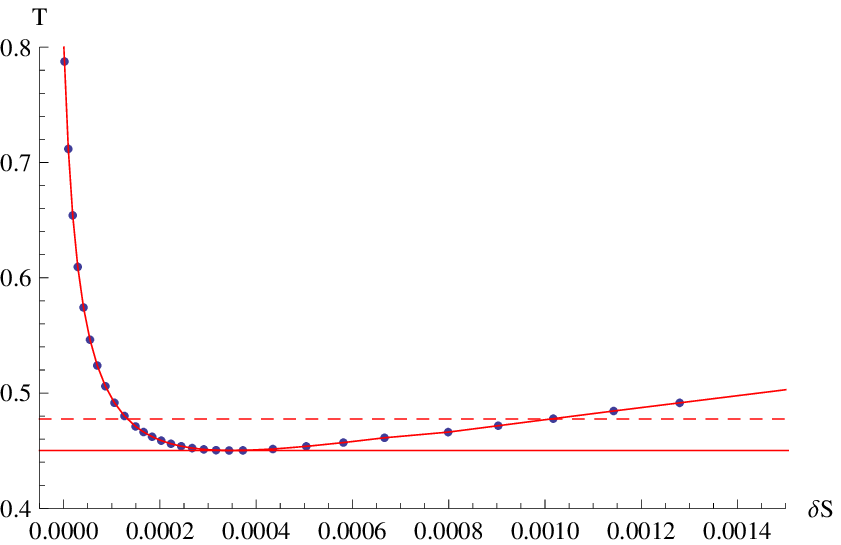}  }
\subfigure[$Q=\frac{\pi -1}{6 \sqrt{5}}$]{
\includegraphics[scale=0.75]{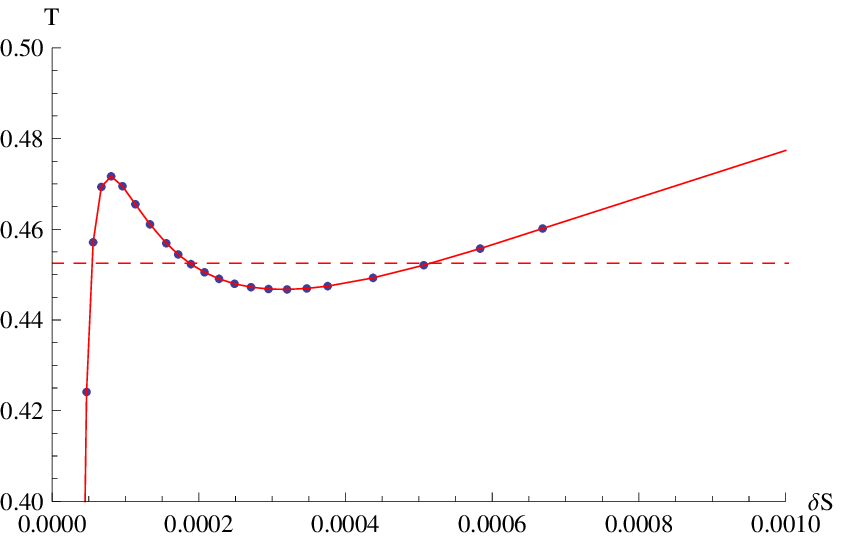}
}
\subfigure[$Q=\frac{\pi }{6 \sqrt{5}}$]{
\includegraphics[scale=0.75]{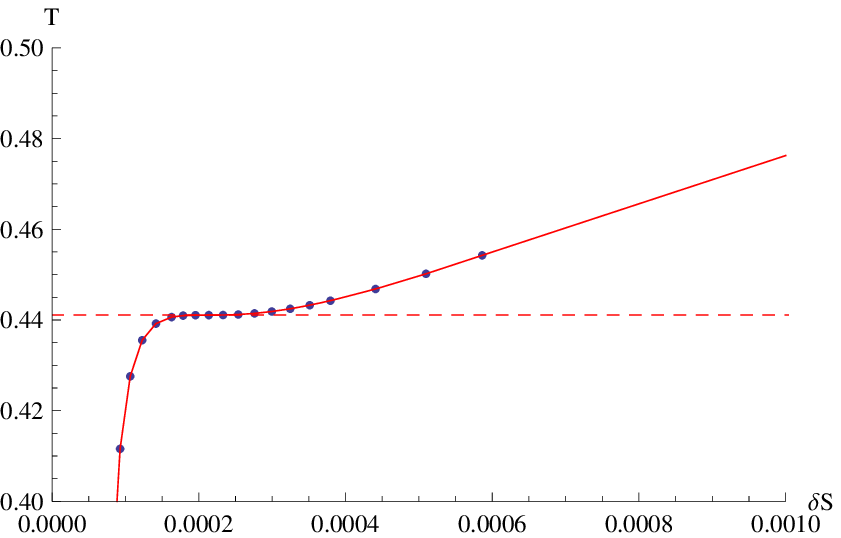}
}
\subfigure[$Q=\frac{\pi +1}{6 \sqrt{5}}$]{
\includegraphics[scale=0.75]{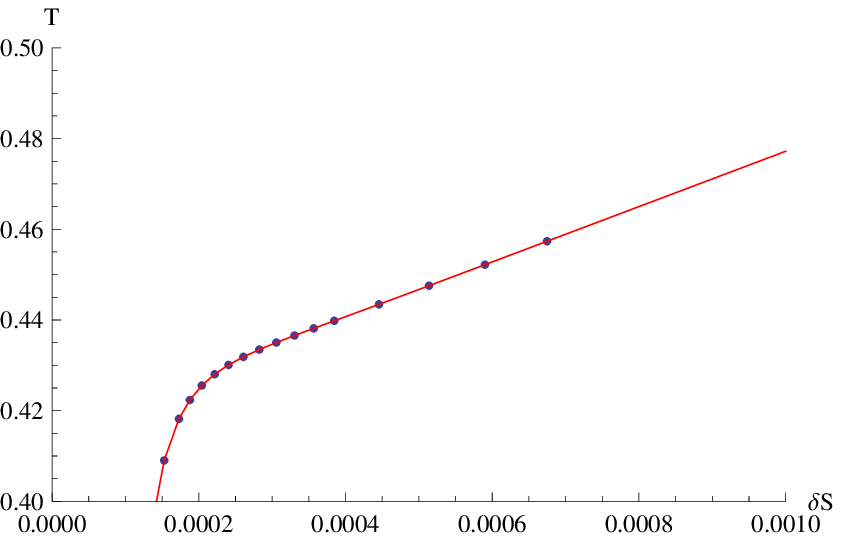}
}
\caption{\small Relation between the entanglement entropy  and temperature  in the fixed charge ensemble for differen charges at $\phi_0=0.2$.  The red solid line corresponds to the location of the minimum temperature $T_0$,  the  dashed
  lines  in (a), (b), (c) correspond individually to the locations of Hawking-Page phase transition $T_1$, first order phase transition $T_{\star}$, and second order phase transition $T_c$.} \label{fig10}
\end{figure}
 Figure \ref{fig9} and  Figure \ref{fig10}, respectively. As one can see, it exhibits a similar behavior as the thermal entropy. To be more precise, we would like to check the equal area law with the  following equation
\begin{eqnarray}
A_1\equiv\int_{\delta S_1}^{\delta S_3} T(\delta S) d\delta S= T_{\star} (\delta S_3-\delta S_1)\equiv A_3,\label{arealaw1}
 \end{eqnarray}
in which  $T(\delta S)$  is an Interpolating Function obtained from the numeric result and $\delta S_1$,  $\delta S_3$ are the smallest and largest roots of the equation $T(\delta S)=T_{\star}$.
For different  $\phi_0s$, the results of  $\delta S_1$,  $\delta S_3$  and $A_1$,  $A_3$ are listed in Table
\ref{tab2}.
\begin{table}
\begin{center}\begin{tabular}{c|c|c}
 %\MC{3}{c}{\text{caption}}\\[5pt]
 \hline
% & \multicolumn{3}{c||}{MGCDM}   & \multicolumn{3}{c}{$\Lambda$CDM}  \\ \hline
%                             &        MGCDM        &                  &             &      $\Lambda$CDM    &                   & \\ \hline
% \MC{3}{|c|c|}{\ZZ{-8pt}{15pt}\hfill\normalsize   \hfill  \hfill\normalsize MGCDM     \hfill\normalsize $\Lambda$CDM  }\\ \hline
% \ZZ{-6pt}{22pt}
 $\phi_0=0.14$ &                                          $\phi_0=0.2$         \\ \hline
                   $T_{\star}$ =0.4526              &    $T_{\star}$ =0.4526       \\    \hline
$\delta S_1$=0.0000155355$\mid$  $\delta S_3$=0.000123195&   $\delta S_1$=0.000186918$\mid$  $\delta S_3$=0.000517552           \\\hline
$A_1$=0.0000487431$\mid$  $A_3$=0.0000487266&   $A_1$=0.000148384$\mid$  $A_3$=0.000149645
  \\ \hline
\end{tabular}
\end{center}
\caption{Check of the equal area law in the $T-\delta S$ plane for different $\phi_0$.}\label{tab2}
\end{table}
It is obvious that $A_1$ nearly equals   $A_3$  regardless of the choice of $\phi_0$. That is, the equal area law is also valid for the entanglement entropy.

\begin{figure}
\centering
\subfigure{
\includegraphics[scale=0.75]{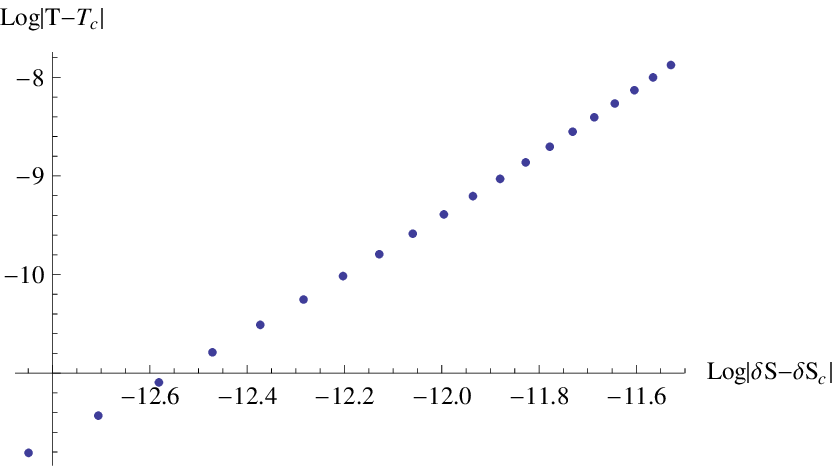}
 }
 \subfigure{
\includegraphics[scale=0.75]{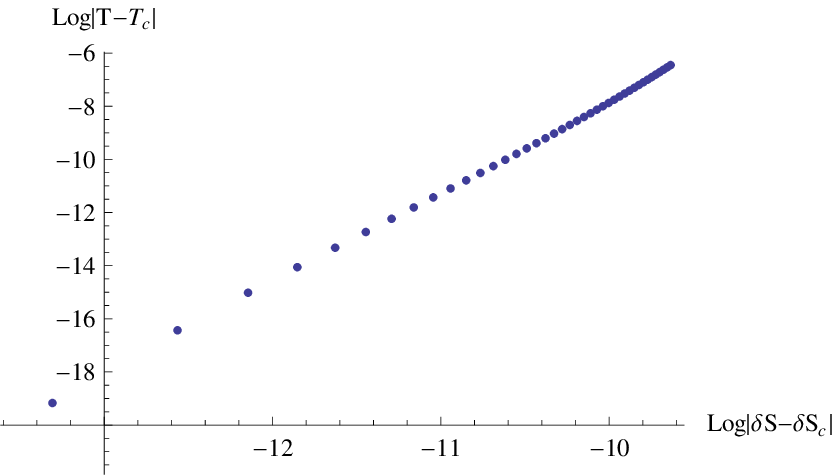}
 }
 \caption{\small Relation between  $\log\mid T-T_c\mid$ and $\log\mid\delta S-\delta S_c\mid $ near the critical point of second order phase transition for different $\phi_0$} \label{fig11}
\end{figure}

To get the critical exponent of second order phase transition of entanglement entropy,
we  should find the slope of a linear function represented by
$ \log\mid T-T_c\mid$ and $\log\mid\delta S-\delta S_c\mid  $,
in which
 $S_c$ is the critical entropy obtained numerically by the equation $T(\delta S)=T_c$. The numeric results for different $\phi_0$ are plotted in Figure
  \ref{fig11}.
 The results for these curves can be further fitted as
\begin{equation}
\log\mid T-T_c\mid=\begin{cases}
26.653 + 3.00107  \log\mid\delta S-\delta S_c\mid,&  $for$ ~\phi_0=0.14\\
21.2674 + 2.92789  \log\mid\delta S-\delta S_c\mid, &~$for$~\phi_0=0.2\\
\end{cases}
\end{equation}
One can see that the slope is always about 3 for different $\phi_0$.  So we can conclude that the entanglement entropy also has the same second order phase transition as  the thermal entropy.

%======================figure1====================

\section{Concluding remarks}
Investigation on the phase transition of the black holes is important and necessary. On one hand, it is helpful for us to understand the structure and nature of the space time.  On the other hand, it  may uncover some phase transitions of the realistic  physics in the conformal field theory according to the AdS/CFT correspondence. It is well-known now that the Hawking-Page phase transition in the gravity system is dual to the confinement/deconfinement phase transition, and the phase transition of a scalar field is dual to the superconductivity phase transition in the dual conformal field theory.

In this paper, we investigated the  van
der Waals-like  phase transition in the  framework of holography so that we can  explore
whether there is a  realistic similar phase transition  in
physics.
Taking  the 5-dimensional  Reissner-Nordstr\"{o}m-AdS  black hole as the gravity background, we  investigated  the phase structure of the two point correlation function, Wilson loop, and  holographic entanglement entropy. For all the non-local observables, we observed that the black hole undergos a van
der Waals-like  phase transition.
 This conclusion is reinforced by the investigation of the equal area law  and critical exponent of the analogous heat capacity¡ê¡§ in which we found that the equal area law is valid always and the critical exponent of the heat capacity coincides with that of the mean field theory   regardless of the  size of the boundary region.
 In addition, we found the black hole  undergos a Hawking-Page phase transition before the  van
der Waals-like  phase transition for all the non-local observables. We also obtained the minimum temperature and Hawking-Page phase transition temperature.
  Our investigation thus provides a complete picture depicting the phase transition of charged AdS  black hole in the
  framework of holography.

\section*{Acknowledgements}

We would like to thank Rong-Gen Cai for his discussions. This work is supported  by the National
Natural Science Foundation of China (Grant Nos. 11405016, 11575270), China Postdoctoral Science Foundation (Grant No. 2016M590138), Natural Science
Foundation of  Education Committee of Chongqing (Grant No. KJ1500530), and Basic Research Project of Science and Technology Committee of Chongqing(Grant No. cstc2016jcyja0364).

\end{document}